\documentclass[pra,twocolumn,floatfix,showpacs]{revtex4}
\usepackage{graphicx,amsmath}
\newcommand{\bra}[1]{\langle#1|}          
\newcommand{\ket}[1]{|#1\rangle}             
\newcommand{\inprod}[2]{\langle#1|#2\rangle} 
\begin{document} 
\title{Continuous variable quantum teleportation with a finite-basis entanglement resource}

\author{S. I. J. Kurzeja and A. S Parkins}
\affiliation{
Department of Physics, University of Auckland, Auckland,
New Zealand
}
\date{\today}
\begin{abstract}
Entanglement is a crucial resource in quantum information theory. We investigate the use of different forms of entangled states in continuous variable quantum teleportation, specifically the use of a finite-basis entanglement resource. We also consider the continuous variable teleportation of finite-basis states, such as qubits, and present results that point to the possibility of an efficient conditional scheme for continuous variable teleportation of such states with near-unit fidelity using finite-basis entanglement. 
\end{abstract}
\maketitle


Quantum teleportation was first proposed by Bennett {\em et al.} \cite{bennett} primarily for
quantum states of finite-dimensional systems (e.g., two-state
systems). Such ``finite-basis'' teleportation requires joint
measurement in the Bell-operator basis of the system whose state is to be teleported and one component
of an entangled bipartite system. This necessitates nonlinear
quantum-quantum interactions and the ability to distinguish between
$N^2$ states for an $N$-dimensional system
\cite{nogovaid,nogolutk}. These are difficult and demanding
requirements which have limited the efficiency (or fidelity) of teleportation achieved to date
in experimental demonstrations involving the polarization states of
single photons \cite{expqubit,expqubit1,expqubit2}.

The procedure for teleportation has also been extended to continuous
variable (i.e., infinite-dimensional) systems \cite{vaidman94, nogovaid} and a specific proposal for
quadrature phase amplitudes of the electromagnetic field was put
forward \cite{cts_braun} and then demonstrated experimentally
\cite{exp_furusawa}. In this case, one requires (two-mode) squeezed states, linear mixers
(i.e., beamsplitters), and quadrature phase measurements, all of which
are well-established in the field of quantum optics and can be produced 
or implemented with reasonably high efficiency. Of course, any quantum state can in principle be teleported using the
continuous variable quantum teleportation (CVQT) scheme, so it is
clear that there could be some advantage to using this scheme for the
teleportation of finite-basis states (such as qubit states) \cite{cvqt_single_photon}, in comparison 
to the original finite-basis teleportation scheme of Bennett {\em et al.} \cite{bennett}. The states of finite-dimensional systems could
be encoded straightforwardly onto number (Fock) states of modes of the electromagnetic field.
However, perfect CVQT requires perfect entanglement between EPR
(Einstein-Podolsky-Rosen) correlated modes, or, in other words, perfect
squeezing. Unfortunately, this requires infinite energy, which can be seen from the number state expansion for the two-mode
squeezed vacuum (2MSV) state,
\begin{align}
   \ket{\Psi_{\text{2MSV}}}_{23}=\sqrt{1-\lambda^2}\sum_{n=0}^{\infty}\lambda^n\ket{n}_2\ket{n}_3
    \label{eq:fock2msv}
\end{align}
where $\lambda$, the squeezing parameter, approaches one in the limit
of perfect squeezing, and the mean excitation in each of the modes (labelled 2 and 3) is given by $\lambda^2/(1-\lambda^2)$.
Nevertheless, non-maximally-entangled 2MSV states
($\lambda < 1$), as produced, for example, by nondegenerate optical
parametric amplification, can still facilitate CVQT with a fidelity of
teleportation higher than can be achieved by classical means only
\cite{cts_braun,exp_furusawa}.  


In this paper, we consider the teleportation procedure of
Braunstein and Kimble \cite{cts_braun} but with an alternative
truncated entanglement resource. There has been some investigation
into the improvement of CVQT by manipulating the 2MSV entanglement
resource with conditional measurements \cite{opatrny_epp,cochrane_epp}, and into the use of truncated entanglement resources in a CVQT scheme involving number-difference and phase-sum (as opposed to quadrature phase) measurements \cite{cochrane_np}. Here, we consider the case in which
the entanglement resource, perhaps prepared also through manipulation
of 2MSV states, or possibly by some other means (see below), takes
the form of a maximally entangled state between $N$-dimensional
subspaces (MEND state) of the (infinite-dimensional) Hilbert
spaces of the entangled oscillator modes, i.e., 
\begin{align}
  \ket{\Psi_{\text{MEND}}}_{23}=\frac{1}{\sqrt{N}}\sum_{k=0}^{N-1}\ket{k}_2\otimes\ket{k}_3\;,
  \label{eq:maxent}
\end{align}
with $\{\ket{k}\}$ the Fock basis. We show that with this entanglement
resource the teleportation fidelity for given quadrature
(``Bell-state'') measurement results is directly proportional to the
probability of obtaining those results and that, for sufficiently
large $N$, there exists a broad and uniform region of measurement
results over which the fidelity is essentially equal to one. This
suggests the possibility of a highly efficient conditional
teleportation scheme, particularly for quantum states in lower
dimensional subspaces (e.g. qubits).

As far as preparation of states of the form (\ref{eq:maxent}) is
concerned, there has been a proposal based on an entanglement
purification scheme for 2MSV states, specifically for optical
systems \cite{cts_epp}. This technique can also be mapped
onto other harmonic oscillator systems, such as the quantized motional
modes of trapped atoms \cite{tratomEPP}.

In addition, entangled collective spin states of atomic ensembles in the form
(\ref{eq:fock2msv}) have recently been experimentally created
\cite{spinlonglivedentanglement}, with a technique that should also
enable the preparation of states of the form (\ref{eq:maxent}) for
finite ensembles \cite{spinpairwiseentanglement,berrysandersswapping}. Such entangled spin
states can in principle be mapped onto light fields \cite{sqlighttospinsqatoms}, or Bell-state measurements as required for quantum teleportation can be implemented via dispersive coupling to laser fields \cite{spinlonglivedentanglement}.
 

To begin our analysis, let us first outline the teleportation protocol
of Braunstein and Kimble \cite{cts_braun}. Let the input quantum state
be in system 1, while the entanglement resource is shared between
systems 2 and 3, where system 3 is the final target system. The usual
characters Alice (system 2) and Bob (system 3) share the entangled
resource, and Alice is given the desired unknown quantum state,
$\ket{\psi_{\text{in}}}$, to be transferred to Bob. Alice combines this
state with her portion of the entangled resource and makes joint
measurements of the position difference,
$\hat{x}^{-}_{12}=\hat{x}_{1}-\hat{x}_2$, and momentum sum $
\hat{p}_{12}^{+}=\hat{p}_{1}+\hat{p}_2$, observables (which commute).
She then communicates (classically) the results of these
measurements to Bob, who makes an appropriate coherent displacement of
his mode to complete the process.

To formulate the teleportation process mathematically, we use the transfer operator introduced in \cite{teleport_operator}, which relates the teleported state $\ket{\psi_{\text{out}}}$ to the input state as
\begin{align}
  \ket{\psi_{\text{out}}}=\hat{T}\ket{\psi_{\text{in}}}\;.
\end{align}
For perfect teleportation the transfer operator is simply the identity operator. The output state is not normalized, and its scalar product gives the probability of measuring a particular value $\beta=x^-_{12} + ip^+_{12}$,
\begin{align}
  P(\beta)=\inprod{\psi_{\text{out}}}{\psi_{\text{out}}}=\bra{\psi_{\text{in}}}\hat{T}^\dagger\hat{T}\ket{\psi_{\text{in}}}\;.
\label{eq:probT}
\end{align}
The teleportation fidelity is given by the overlap between the input and output state,
\begin{align}
  F(\beta)&=\frac{1}{P(\beta)}\left|\inprod{\psi_{\text{in}}}{\psi_{\text{out}}}\right|^2=\frac{1}{P(\beta)}\left|\bra{\psi_{\text{in}}}\hat{T}\ket{\psi_{\text{in}}}\right|^2\;.
\label{eq:fidT} 
\end{align}
The average fidelity $F_{\text{av}}$ is given by
\begin{align}
  F_{\text{av}}=\int d^2\beta P(\beta)F(\beta)=\int d^2\beta\left|\bra{\psi_{\text{in}}}\hat{T}\ket{\psi_{\text{in}}}\right|^2\;.
\end{align}
Hence, the transfer operator holds all information about the
teleportation process. For the 2MSV quantum channel (with unit gain on Bob's displacement operator \cite{polkralphgain,hofmanngain}) the transfer operator is given by
\begin{align}
\hat{T}_{\text{2MSV}}&= \sqrt{\frac{1-\lambda^2}{\pi}} \sum_{n=0}^\infty \lambda^n \hat{D}(\beta)\ket{n}\bra{n}\hat{D}(-\beta)\;.
\label{eq:2msv_cts_op}
\end{align}
For the MEND quantum channel (and unit gain) the transfer operator is,
\begin{align}
  \hat{T}_{\text{MEND}}=\frac{1}{\sqrt{\pi N}}\sum_{n=0}^{N-1} \hat{D}(\beta)\ket{n}\bra{n}\hat{D}(-\beta) \;.
  \label{eq:mend_cts_op}
\end{align}
For a measurement of $\beta =0$, the teleportation operator is
essentially the identity operator in a finite basis, in which case we
have perfect teleportation of states which involve $N-1$ excitations or less.

The transfer operator for the MEND quantum channel given in equation (\ref{eq:mend_cts_op}) has the following property,
\begin{align}
\hat{T}_{\text{MEND}}^\dagger\hat{T}_{\text{MEND}}=\frac{1}{\sqrt{\pi N}}\hat{T}_{\text{MEND}}\;.
\end{align}
Therefore, the probability distribution shown in equation (\ref{eq:probT}) may be written as
\begin{align}
  P_{\text{MEND}}(\beta)=\frac{1}{\sqrt{\pi N}}\bra{\psi_{\text{in}}}\hat{T}_{\text{MEND}}\ket{\psi_{\text{in}}}\;.
\end{align}
The fidelity distribution defined in (\ref{eq:fidT}) can then be written in terms of $P_{\text{MEND}}(\beta)$ as
\begin{align}
  F_{\text{MEND}}(\beta)=\pi N P_{\text{MEND}}(\beta)\;,
  \label{eq:fidprobN}
\end{align} 
which shows that, independent of the state we are teleporting, the fidelity and probability distribution have the
same functional form, differing only by the constant $\pi N$. It shows a direct connection between the amount of
information gained from the measurement and the disturbance to the
teleportation process. Note, $F(\beta)$ must be bounded
by the set of real numbers $[0,1]$, which means $P(\beta)$ must 
be bounded by the set $[0,1/(\pi N)]$. Since $P(\beta)$ will have non-zero values and must be normalized, it is of compact support, that is, bounded
and closed for some region. It follows that $F(\beta)$ will also be of compact support and as $N$ is increased the maximum probability is
reduced, and consequently $P(\beta)$ becomes
broader, as does $F(\beta)$. These comments are general but the precise nature of the
distributions depends on the type of state being
teleported, so we now consider several popular examples.

\begin{figure}[t]
\includegraphics[width=8cm]{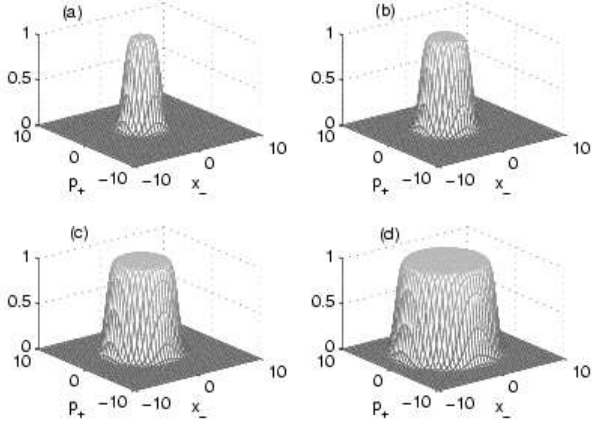}
\caption{Fidelity distribution for teleporting a coherent state with
  coherent amplitude $\alpha = 1.5i$ using a MEND quantum channel for
  different truncation number (a)$N=6$, (b) $N=11$, (c) $N=21$, and (d) $N=41$.}
\label{fig:cohrmend}
\end{figure}

Suppose the input state to teleport is a coherent state,
$\ket{\psi_{\text{in}}}=\ket{\alpha}\;.$
For the 2MSV quantum channel, using the transfer operator in equation (\ref{eq:2msv_cts_op}) we obtain the usual Gaussian fidelity distribution 
\begin{align}
F(\beta)=e^{-(1-\lambda)^2|\alpha-\beta|^2}\;,
\label{eq:fidcohr2msv}
\end{align}
with the probability for measuring $\beta$ given by
\begin{align}
P(\beta)=\frac{1-\lambda^2}{\pi}e^{-(1-\lambda^2)|\alpha-\beta|^2}\;.
\end{align}
For the MEND quantum channel, the fidelity distribution found using the transfer operator (\ref{eq:mend_cts_op}) is
\begin{align}
F(\beta)&=e^{-|\alpha-\beta|^2}\sum_{n=0}^{N-1}\frac{|\alpha-\beta|^{2n}}{n!}\;,\notag\\
&=1 -\exp(-|\alpha-\beta|^2)\sum_{n=N}^{\infty}\frac{|\alpha-\beta|^{2n}}{n!}
\label{eq:fidcohrmend}
\end{align}
with, as shown before, the probability of measuring $\beta$ given by
$P(\beta) = 1/(\pi N)F(\beta)$. It follows from (\ref{eq:fidcohrmend})
that high fidelities close to one occur when the Taylor expansion
to order $N$ of $\exp(|\alpha-\beta|^2)$ is a good approximation to
$\exp(|\alpha-\beta|^2)$. Hence, for sufficiently large $N$, the fidelity
distribution has the significant feature of being flat and essentially
equal to one over the region where the residual of the Taylor expansion, 
$\sum_{n=N}^{\infty}\frac{|\alpha-\beta|^{2n}}{n!}$ is small. This
is shown in Fig. \ref{fig:cohrmend} for different values of
truncation number $N$ in the MEND quantum channel. In contrast, with the 2MSV resource the fidelity is only equal to one
for $\beta = \alpha$ and drops off rapidly for $\beta \neq \alpha$ (
unless $\lambda$ is close to one).

\begin{figure}[t]
\includegraphics[width=8cm]{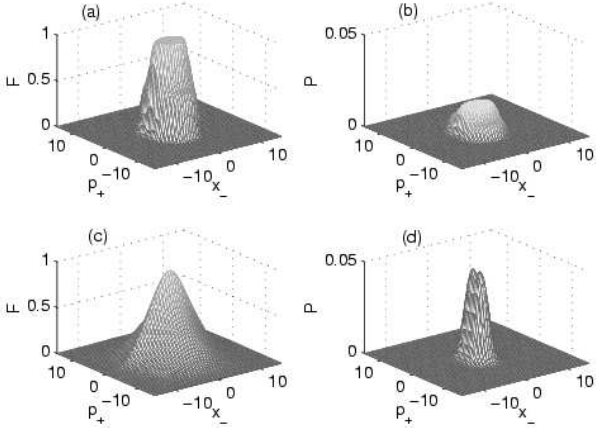}
\caption{Fidelity and probability distribution for teleporting a ``Schr{\"o}dinger Cat'' state with $\alpha=1.5i$ comparing (a),(b) the MEND quantum channel with $N=21$ and (c), (d) the 2MSV quantum channel with $\lambda = 0.85$.}
\label{fig:cat}
\end{figure}
\begin{figure}[htpb]
\includegraphics[width=8cm]{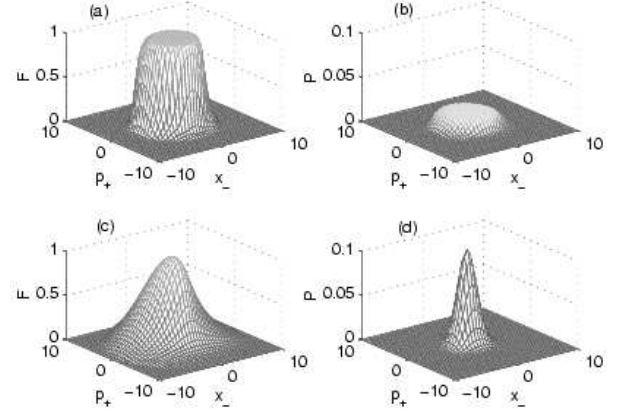}
\caption{Teleporting the qubit state $\ket{\psi_{\text{in}}}=1/\sqrt{2}(\ket{0}+\ket{1})$. (a), (b) show the
  fidelity and probability distributions for the MEND
  quantum channel with $N=21$. (c) and (d) show the fidelity and
  probability distributions for the 2MSV with $\lambda=0.8$}
\label{fig:qubit}
\end{figure}

Another standard example in the context of CVQT is the
``Schr{\"o}dinger Cat'' state, $\ket{\psi_{\text{in}}} =
{\cal N}_\alpha(\ket{\alpha}+ \ket{-\alpha})$ (${\cal N}_\alpha =
[2+2e^{-2|\alpha|^2}]^{-\frac{1}{2}}$)
\cite{cts_braun,opatrny_epp,cochrane_epp}. The fidelity and
probability distributions for this state are shown in
Fig. \ref{fig:cat} for both the MEND ($N=21$) and 2MSV ($\lambda =
0.85$) entanglement resources. Again, we observe a distinct ``flat-top'' region for the
MEND resource where the fidelity is equal to one, as opposed to
isolated maxima for the 2MSV resource.


Now let us suppose that the input state is a general finite-basis pure
state encoded on harmonic oscillators,
\begin{align}
\ket{\psi_{\text{in}}}=\sum_{m=0}^{M}c_m\ket{m}\;,
\end{align}
and let us also assume a general entangled state resource 
\begin{align}
  \ket{\psi_{\text{EPR}}}_{23}=\sum_{n=0}^{\infty}d_n\ket{n}_2\ket{n}_3\;,
\end{align}
which has a transfer operator given by,
\begin{align}
\hat{T} = \frac{1}{\sqrt{\pi}}\sum_{n=0}^\infty d_n\hat{D}(\beta)\ket{n}\bra{n}\hat{D}(-\beta) \;.
\end{align}
The action of the displacement operator on a Fock state can be expressed as \cite{2squeezed}
\begin{align}
  \hat{D}(\beta)\ket{n}=e^{-\frac{|\beta|^2}{2}}\sum_{k=0}^{\infty}\sqrt{\frac{n!}{k!}}L_n^{k-n}(|\beta|^2)\beta^{k-n}\ket{k}\;,
\label{eq:displacefock}
\end{align}
where $L_n^{\alpha}$ are the generalized Laguerre polynomials
\cite{mathlag}. The probability distribution is then
\begin{gather}
P(\beta)=e^{-|\beta|^2}G(\beta,\{|d_n|\})\;,
\end{gather}
and the fidelity distribution becomes,
\begin{gather}
F(\beta)=e^{-|\beta|^2}G(\beta,\{d_n\})/G(\beta,\{|d_n|^2\})\;,
\end{gather}
where
\begin{gather}
  G(\beta,\{f_n\})=e^{-|\beta|^2}\times\notag\\
  \sum_{m,l=0}^{M}\sum_{n=0}^{\infty}c_{l}^*c_{m}f_n\frac{n!}{\sqrt{m!l!}}L_n^{m-n}(|\beta|^2)L_n^{l-n}(|\beta|^2)\frac{\beta^{m}\beta^{l}}{|\beta|^{2n}}\;.
  \label{eq:G}
\end{gather}

Fig. \ref{fig:qubit} shows the results for the most fundamental case of a finite-basis state, the
qubit, defined as
\begin{gather}
\ket{\psi_{\text{in}}}=a\ket{0}+b\ket{1},\quad \;a^2 + b^2 = 1\;.
\end{gather}

The same general features are exhibited as for the previous
examples. Despite the broader range of high fidelities for the MEND
resource, an important point to note in the comparison between the
MEND and the 2MSV resources is that the {\em average} teleportation
fidelity is actually very similar for the case shown -- 0.86 for the
2MSV resource with $\lambda = 0.8$, and 0.85 for the MEND resource
with $N=21$. This is because the probability distribution for the 2MSV
resource is narrower than its fidelity distribution, thus giving more
weight to the central regions of higher fidelity.

However, the significant feature that we wish to emphasize is that for
the MEND resource with $N=21$ there is a $48\%$ chance of obtaining
Bell-state measurement results for which the corresponding fidelity of
teleportation is {\em greater} than 0.99. For the 2MSV resource, the
{\em maximum} fidelity, occurring only at the peak of the distribution
shown in Fig. \ref{fig:qubit}(c), is 0.99.

\begin{figure}[t]
\includegraphics[width=5cm]{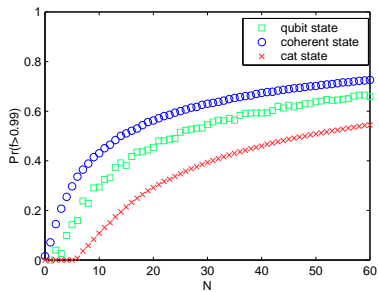}
\caption{The probability of obtaining a fidelity greater than 0.99
  against truncation number $N$ of the MEND quantum channel for
  coherent state, cat state, and qubit inputs.}
\label{fig:selective}
\end{figure}

This points to the possibility of a very efficient {\em conditional}
teleportation scheme using the MEND resource, where, in particular,
the conditioning is upon obtaining Bell-state measurement results
that lie within the region for which the fidelity of teleportation is
known to be very close to one. That is, given some knowledge of the
general class of input state (e.g. coherent state, qubit) and of the
truncation number $N$ of the MEND state, one could select out, or
choose to pursue, only those events where Alice's measurements fall
within the (broad) region that gives near-perfect teleportation. The
potential efficiency of such a conditional teleportation scheme is
characterized in Fig \ref{fig:selective}, where we plot, as a function
of $N$ (and for each of the examples considered), the probability of
Alice obtaining a measurement result for which the teleportation
operation, upon completion, occurs with a fidelity greater than
0.99. As one can see, this probability can be substantial for only
moderately large values of $N$.

In conclusion, the maximally entangled state in a finite-basis subspace of the full infinite-dimensional Hilbert space gives a direct proportionality between the Bell-state measurement probability and the fidelity of teleportation. This relationship constrains the form of the distribution of fidelities over the entire measurement domain to one which has a very flat region of unit fidelity followed by a quick cut off to zero fidelity. Using the measurement result, one can in principle post-select or pursue only those teleportation events which occur in the flat region of high fidelity to implement an efficient conditional teleportation scheme.

The authors acknowledge support from the Marsden Fund of the Royal Society of New Zealand.

\end{document}